\documentclass[showpacs,amsmath,amssymb,prb,
               superscriptaddress,
               twocolumn
              ]{revtex4}
\usepackage[T1]{fontenc}
\usepackage{times}
\usepackage{graphicx}
\usepackage{wasysym}
\usepackage{bm}

\begin{document}

\title{Limits on the amplification of evanescent waves
       of left-handed materials}

\author{Th.~Koschny}
\affiliation{Ames Laboratory and Dept.~of Phys.~and Astronomy,
             Iowa State University, Ames, Iowa 50011, U.S.A.}
\affiliation{Institute of Electronic Structure and Laser, FORTH,
             71110 Heraklion, Crete, Greece}

\author{R.~Moussa}
\affiliation{Ames Laboratory and Dept.~of Phys.~and Astronomy,
             Iowa State University, Ames, Iowa 50011, U.S.A.}

\author{C.~M.~Soukoulis}
\affiliation{Ames Laboratory and Dept.~of Phys.~and Astronomy,
             Iowa State University, Ames, Iowa 50011, U.S.A.}
\affiliation{Institute of Electronic Structure and Laser, FORTH,
             71110 Heraklion, Crete, Greece}

\date{\today}

\begin{abstract}
We investigate the transfer function of the discretized perfect lens in
finite-difference time-domain (FDTD) and transfer matrix (TMM) simulations; 
the latter allow to eliminate the problems associated with the explicit time
dependence in FDTD simulations.
We argue that the peak observed in the FDTD transfer function 
near the maximum parallel momentum $k_{\|,\mathrm{max}}$ 
is due to finite time artifacts.
We also find the finite discretization mesh acts like imaginary
deviations from $\mu=\varepsilon=-1$ and leads to a cross-over
in the transfer function from constance to exponential decay around 
$k_{\|,\mathrm{max}}$ limiting the attainable super-resolution.
We propose a simple qualitative model to describe the impact of the
discretization. 
$k_{\|,\mathrm{max}}$ is found to depend logarithmically on the mesh constant
in qualitative agreement with the TMM simulations.
\end{abstract}



\maketitle

The ability of the left-handed finite slab with a homogeneous permeability 
$\mu=-1$ and permittivity $\varepsilon=-1$ to form a perfect lens (PL) has 
received much attention since first proposed by Pendry\cite{Pendry00}.
Such a slab does not only compensate the phase of the
propagating waves emanating from a point source to form a focus on the
opposite side of the slab. It also amplifies the evanescent waves which decay 
exponentially in vacuum into exponentially growing solutions inside the slab.
This way all the source amplitudes reemerge in the focus.
The immediate consequence of this behavior is that the resolution of the 
image may overcome the diffraction limit.
Soon after, it was realized\cite{Ramakrishna02,Smith03} 
that the restoration of the evanescent waves by the PL is exceptionally 
sensitive to small deviations from $\mu=\varepsilon=-1$. 
The transfer function of the PL, defined by the amplitude ratio of a plane 
wave component at the focus and the source, is, in the ideal case, unity 
for all $\omega$ and $k_\|$ up to infinity.
For the near-perfect lens it exposes an order of unity ($o(1)$) behavior 
at small parallel momenta $k_\|$ which turns into exponential decay 
$\sim e^{-k_\|d}$ for large $k_\|$. 
The cross-over between $o(1)$ behavior and exponential decay for a 
given PL defines a maximum parallel momentum $k_{\|,\mathrm{max}}$
which qualitatively constitutes the highest evanescent wave still
restored by the PL, hence, defines the maximum attainable sub-wavelength
resolution $\Delta x_\mathrm{min}\sim 2\pi/k_{\|,\mathrm{max}}$. 
For small deviations $\mu=\varepsilon=-1+\gamma$ with 
$\gamma\in {\bf C},\ |\gamma|\ll 1$ from the ideal PL 
a logarithmic dependence $k_{\|,\mathrm{max}}\,d\sim -\log\, |\gamma|$ 
of the cross-over momentum has been found\cite{Smith03}.
Here and throughout the paper we employ a dimensionless formulation 
measuring all lengths in units of the linear size $L$ of the unit cell and
all frequencies in  units of the vacuum speed of light divided by $L$. 
In particular, this renders the dimensionless vacuum speed of light $c=1$
and wavelength $\lambda=2\pi/\omega$.

Almost all numerical investigations of the PL's imaging properties 
deploy finite-difference time domain (FDTD) simulation using a time and space 
discretized version of the Maxwell equations.
After a few contradictory publications\cite{Ziolkowski01,Fang03,Cummer03}, 
Rao and Ong\cite{Rao03a,Rao03b} 
established the amplification of the evanescent waves inside the LH material
slab and the cross-over behavior in the transfer function numerically for 
the FDTD method.
They also observed the occurrence of surface plasmons, ie.\@ local field 
enhancement, at the first interface for the slightly lossy PL.
The FDTD simulations of the PL suffer from explicit time dependence. 
Since the time domain simulations involve a finite time window from the 
"switch-on" to the actual measurement of the fields, the results
are obtained as a superposition of a finite width $\omega$-distribution 
around the target frequency $\omega_0$ which narrows as the simulation 
time increases. 
The corresponding transfer function $t_\mathrm{FDTD}(\omega_0)$ differs
considerably from the stationary (frequency-domain for a single frequency
$\omega$) transfer function $t_\infty(\omega)$.
In conjunction with the physically always present dispersion 
$\mu(\omega)$, $\varepsilon(\omega)$ of the left-handed material this leads
to a possible coupling to the surface plasmons on both interfaces of the
slab which in turn causes convergence problems in the FDTD.
The FDTD only converges for the near-perfect lens where the surface plasmons
are damped by the small imaginary part in the LH 
slab\cite{Rao03b,GomezSantos03}. 
If we approach the ideal PL the FDTD ceases to converge which renders 
the method unusable.

Due to the existence of surface plasmons at the LH slab the transfer function 
of the near-perfect lens includes poles along the surface plasmon 
dispersion relation\cite{GomezSantos03} which can be approximated by
$-\tanh(k_\|d/2)=-1\pm\gamma$ 
for $k_\|$ well above the propagating modes.
By virtue of the LH materials dispersion relation this essentially real 
$\gamma$ directly translates into a frequency deviation from $\omega_0$. 
The poles approach $\omega_0$ exponentially for growing $k_\|$.
For small $k_\|$ the poles of the surface plasmons are usually outside the
finite width $\omega$-distribution, we find convergence of the FDTD and the
transfer function is dominated by the stationary transfer function 
$t_\infty(\omega)$ at $\omega_0$. 
For large $k_\|$ the poles are exponentially 
damped in all lossy cases and cease to contribute either.
However, for intermediate $k_\|\approx k_{\|,\mathrm{max}}$ 
the surface plasmon poles constitute the principal contribution to 
$t_\mathrm{FDTD}(\omega_0)$. 
This leads to non-convergence of the FDTD due to the emerging 
"beating pattern", modulated by the frequency difference of 
the two surface plasmon branches as explained by 
G\'omez-Santos\cite{GomezSantos03}.
As we emphasize here, this also explains the unexpected peak 
$t_\mathrm{FDTD}(\omega_0)\gg 1$ around $k_{\|,\mathrm{max}}$ 
in the FDTD transfer function 
found by Rao and Ong\cite{Rao03a} and also confirmed by our own FDTD 
simulations. 
The peak originates from the contribution of the diverging 
$t_\infty(\omega_\mathrm{pole})$ to the magnitude of 
$t_\mathrm{FDTD}(\omega_0)$.
Analytically we would expect a monotonous transition from the $o(1)$ behavior
below the cross-over momentum below $k_{\|,\mathrm{max}}$ to exponential
decay above for the near-perfect lens.

In general we are interested in the stationary case transfer function
$t_\infty(k_\|,\omega,d)$ of the PL as this allows us to estimate the imaging
properties. The field components in the focus are given by the field 
components in the source as
\begin{displaymath}
E_\mathrm{focus}(k_\|, t)
\ =\
\int d\omega\
t_\infty(k_\|,\omega)\, E_\mathrm{source}(k_\|)\, g(\omega,\omega_0)\,
e^{i\omega t}
\end{displaymath}
where $g(\omega,\omega_0)$ is the frequency distribution around the frequency
of the point source due to the switch-on of the source and the finite 
observation time window. $E_\mathrm{source}(k_\|)$ and $g(\omega,\omega_0)$
are parameters of the setup and measurement, only $t_\infty(k_\|,\omega)$ is
an intrinsic property of the lens.
In the FDTD this stationary transfer function $t_\infty(k_\|,\omega)$ is only 
accessible via the temporal Fourier transform of the simulation results which
is neither convenient nor especially robust against numerical error.

In contrast, the transfer matrix method (TMM) simulations provide a means 
to directly obtain the stationary transfer function for a single frequency.
We can simulate transmission and reflection amplitudes as well as the spatial
field distribution without the artifacts of finite time simulations.
In order to eliminate the effect of numerical error we employ an 
arbitrary-precision implementation of the TMM described for 
instance in Ref.~\onlinecite{Markos02a}.

\begin{figure}
\centerline{\includegraphics[width=8.cm]{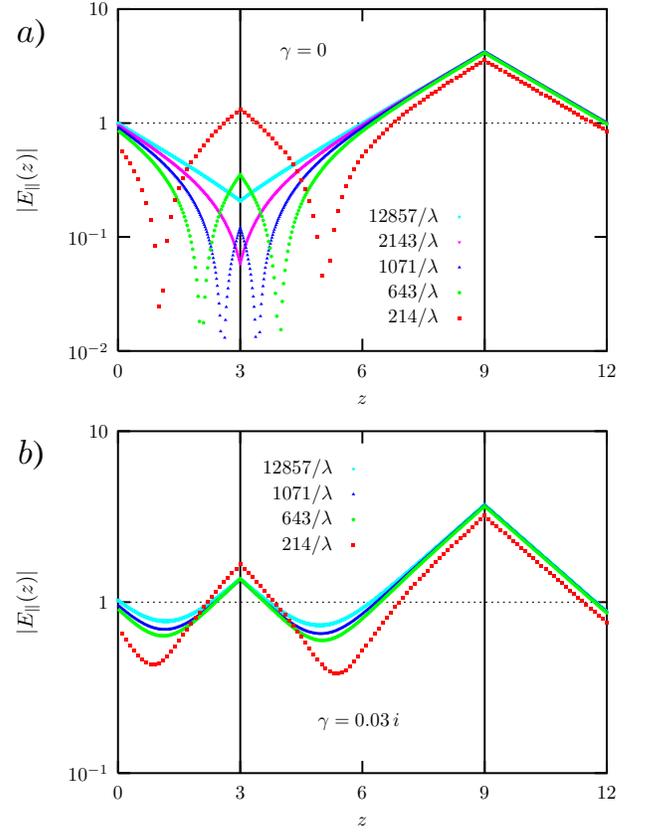}}
 \caption{%
 (Color online)
 The distribution of the electric field in TE mode is shown from source ($z=0$)
 to focus ($z=12$) across the lossless (a) and lossy (b) perfect lens
 for successively finer discretization.
 The left-handed slab between the interfaces at $z=3$ and $z=9$ has
 $\mu=\varepsilon=-1$ for the lossless PL and 
 $\mu=\varepsilon=-1+\gamma$ with $\gamma=0.03\,i$ for the lossy PL.
 We discretized using a uniform cubic mesh with a linear resolution
 ranging from 240 to 12857 mesh points per vacuum-wavelength. 
 }
 \label{fig:tmm-fields}
\end{figure}

Fig.~\ref{fig:tmm-fields}(a) shows the TMM simulated spatial field distribution 
for the parallel component of the E field in TE polarization across a PL for 
one particular evanescent wave component with $k_\|=1.89\,\omega$ and
several spatial discretizations ranging from 240 to 12857 linear mesh points 
per vacuum-wavelength.
The LH slab extends from $z=3$ to $z=9$, which corresponds to a thickness of
$0.28\lambda$. The vertical solid lines indicate the interfaces.
The source is located at $z=0$ and we get an image at $z=12$.
Note that despite the elimination of the explicit time-dependence of the FDTD
simulations we still observe unexpected field enhancement at the first 
interface of the PL.
These artifacts are easily associated with the finite discretization leading 
to reflection of incident evanescent modes at the first interface. Only for
very fine discretization meshes the field distribution approaches the 
analytically expected zig-zag-form featuring a minimum at the first interface.
For coarse discretizations we observe a prominent maximum at the first 
interface accompanied by (almost) zeros of the fields before and after the 
interface indicating a phase shift of the response of the interface.
In Fig.~\ref{fig:tmm-fields}(b) we show the corresponding field distributions
for the lossy PL where a small imaginary part $\gamma=0.03\,i$ is added
to the permeability and permittivity of the LH slab.
Again we observe the field enhancement at the first interface, 
but the zeros of the field have disappeared.
However, in this case the behavior is dominated by the losses in the LH slab 
and the dependence on the discretization is much weaker.
The observed dependence on the imaginary part for fixed discretization 
(not shown) confirms previous results 
obtained from FDTD simulations\cite{Rao03a}.

For the lossy PL there is a simple physical explanation for the 
reflection of evanescent
waves and the occurrence of surface modes at the first interface:
For the evanescent waves in vacuum between source and first interface 
${k}_\|$ is real and ${k}_\bot$ purely imaginary.
Whenever the PL involves an (causal) imaginary part or the evanescent solutions
on the right hand side of the slab couple to propagating modes 
or are subject to absorption, electromagnetic field energy is dissipated in
the system. This energy has to be provided by the source and transmitted 
across the vacuum gap before the PL.
Although it is well known that a single evanescent wave cannot transmit 
energy this is not true for a superposition 
$A\,e^{i({\bf k}_\|+{\bf k}_\bot){\bf r}}+
 B\,e^{i({\bf k}_\|-{\bf k}_\bot){\bf r}}$
of incoming and reflected evanescent wave component. 
The general equation for the time-averaged Poynting vector for the TE mode
inside the vacuum slab is
\begin{eqnarray}
\lefteqn{\left<{\bf S}\right>_\mathrm{TE}^{} \ =\ 
\mathrm{Im}(AB^*)\frac{i\,{\bf k}_\bot}{\omega\mu} \ +\
} \\
&&
\frac{{\bf k}_\|}{\omega\mu}
\left(
\frac{|A|^2}{2}e^{-2\mathrm{Im}({\bf k}_\bot {\bf r})}+
\frac{|B|^2}{2}e^{+2\mathrm{Im}({\bf k}_\bot {\bf r})}+\mathrm{Re}(AB^*)
\right)
\nonumber
\end{eqnarray}
and similarly for the TM mode.
Considering the first term it is immediately clear that in order to have an 
energy current normal to the interface across the gap, $\mathrm{Im}(AB^*)$ 
and thus the reflection at the first interface has to be non-zero. 
Note that this even applies to the ideal PL if we have an outgoing energy 
current or dissipation on the right hand side of the lens.

\begin{figure}
\centerline{\includegraphics[width=8.cm]{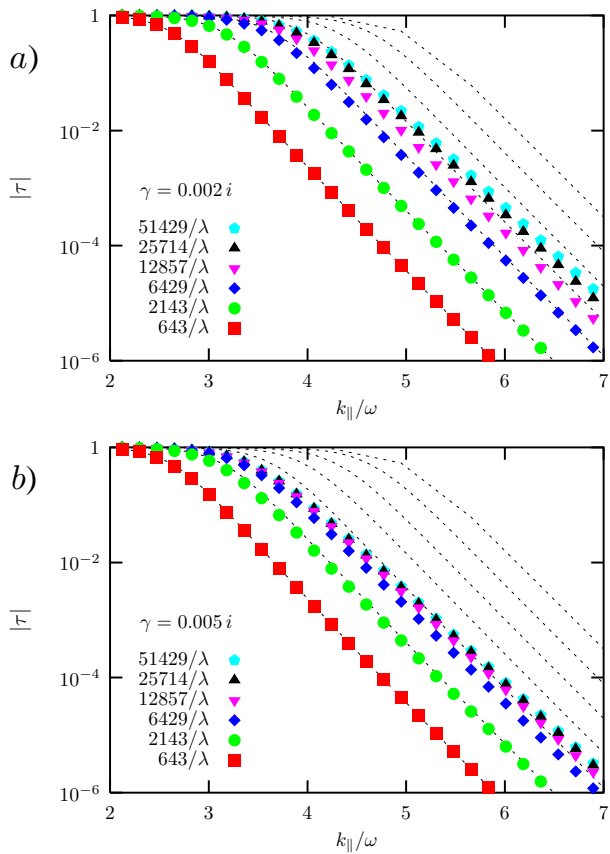}}
 \caption{%
 (Color online)
 The transfer function from source to focus is shown for the lossy PL (symbols)
 with $\mu=\varepsilon=-1+\gamma$ for two different small imaginary parts
 $\gamma=0.002\,i$ (a) and $\gamma=0.005\,i$ (b) and different 
 discretizations.
 The dashed lines show the corresponding transfer function for the lossless PL.
 We discretized using a uniform cubic mesh with a linear resolution
 ranging from 643 to 51429 mesh points per vacuum-wavelength.
 The additional rightmost dashed line corresponds to $102858/\lambda$.
 }
 \label{fig:tmm-t-imag}
\end{figure}

\begin{figure}
\centerline{\includegraphics[width=8.cm]{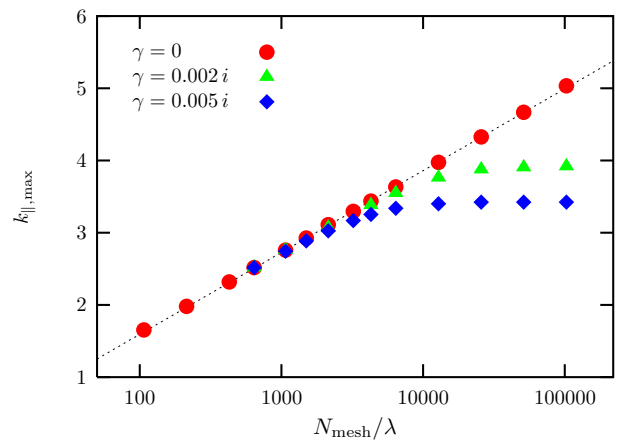}}
 \caption{%
 (Color online)
 The logarithmic scaling of the cross-over parallel momentum
 $k_{\|,\mathrm{max}}$ with the number of linear mesh points per
 vacuum-wavelength is shown for the lossless ($\gamma=0$) and two 
 lossy PLs with $\gamma=0.002\,i$ and $\gamma=0.005\,i$.
 The dotted line is a fit
 $k_{\|,\mathrm{max}} d = (24/27) \log\,(4\lambda/N_\mathrm{mesh})$
 for the lossless PL.
 %
 }
 \label{fig:tmm-t-mesh}
\end{figure}

Now we shall consider the transfer function $t_\infty(k_\|,\omega,d)$ of the
PL with and without losses from source to focus as obtained by the TMM.
In Fig.~\ref{fig:tmm-t-imag} we show the dependence of the transfer function 
for a fixed frequency $\omega=3/10$ on the parallel momentum $k_\|$ for
several spatial discretizations for the lossless PL and two lossy PLs with
imaginary parts $\gamma=0.002\,i$ and $\gamma=0.005\,i$ added to both the
permeability and the permittivity of the LH slab.
Let us first consider the lossless PL represented by the dashed lines in both
panels. For all discretizations there is clear evidence for a cross-over from
$o(1)$ behavior to exponential decay in the $k_\|$ dependence of the transfer
function. The cross-over occurs monotonously without a peak near 
$k_{\|,\mathrm{max}}$ and $k_{\|,\mathrm{max}}$ increases with finer 
spatial discretization.
This indicates that the discretization mesh constant acts like an 
{\em effective} imaginary part in a continuous lossy PL.
For the lossy discretized PL, ie.\@ adding an explicit imaginary part $\gamma$
we observe the same qualitative behavior. 
Here the cross-over is determined by both, discretization and the losses due
to the explicit imaginary parts.
For small $\gamma$ and coarse discretization the behavior of the transfer 
function is entirely dominated by the finite discretization: the lossy PL
virtually coincides with the transfer function for the lossless PL.
For successively finer discretizations $\gamma$ starts to dominate the 
behavior, leading to a saturation of the discretization dependence of 
$k_{\|,\mathrm{max}}$ at a value determined by $\gamma$.
These results show that for a given lossy PL there is always a minimum 
discretization mesh constant where the transfer function "converges", 
ie.\@ becomes independent on the discretization.
For the simulated lossless perfect lens the cross-over in the transfer 
function due to the discretization becomes the primary limiting factor 
for the observation of sub-wavelength resolution. 
In Fig.~\ref{fig:tmm-t-mesh} we show the dependence of the cross-over 
momentum $k_{\|,\mathrm{max}}$ on the discretization for the lossless 
and two lossy $0.28\lambda$-PLs at $\omega=3/10$ 
as extracted from the data presented in Fig.~\ref{fig:tmm-t-imag}.
%
%
It is evident that for the lossless PL over a wide range of discretizations
$k_{\|,\mathrm{max}}$ increases logarithmically with the linear 
number of mesh points per vacuum wavelength. 
For the lossy cases this slow increase saturates at a finite 
$k_{\|,\mathrm{max}}$ which in turn decreases with increasing deviation 
$\gamma$ from the lossless case.
In order to achieve a moderate five-times better resolution than the one 
provided by 
the propagating modes alone for the $d=0.28\lambda$ lossless PL, 
we have to push the discretization to a ridiculously
high value of $10^5$ linear mesh points per vacuum wavelength.
Such discretization mesh densities are easily limited by the available 
computer power.

The effect of the discretization can be qualitatively understood in terms 
of a simple model. 
In the standard discretization of the Maxwell equations the E and H
field components are assigned to the links of two mutually dual 
lattices\cite{Tavlove}.
As a consequence a wave traveling towards the surface of a discretized 
homogeneous slab will first "see" the electric response and approximately 
half a mesh step later the magnetic response (or vice versa, depending on 
the material discretization and definition of the interface).
This can be analytically modeled assuming a continuum lossless PL with 
$\mu=\varepsilon=-1$ to be sandwiched between two thin layers with 
$\mu=-\varepsilon=1$ and $-\mu=\varepsilon=1$, respectively.
The thickness $\delta$ of the surface layers shall be of the order of the 
discretization mesh constant. Now we can derive the leading order 
$\delta$-corrections to the transfer function analytically using the
transfer matrix technique.
We can calculate the total transfer matrix of the left-handed slab 
(${\bf \tau}_{2})$ wrapped in surface layers (${\bf \tau}_{1}, {\bf \tau}_{3}$)
and the two surrounding vacuum slabs (${\bf \tau}_{2})$ as
\begin{equation}
{\bf \tau}_\mathrm{imaging}^{} \ =\
{\bf \tau}_{0}^{}(b)\,
\big[
{\bf \tau}_{3}^{}(\delta)\ {\bf \tau}_{2}^{}(d)\ {\bf \tau}_{1}^{}(\delta)
\big]
\,{\bf \tau}_{0}^{}(a).
\quad
\end{equation}
For the transfer matrix of an homogeneous slab in wave representation
we find 
\begin{displaymath}
 \renewcommand{\arraystretch}{1.3}
 {\bf \tau}_{i}^{}(d)
 \ =\
 \left(\!
  \begin{array}{cc}
    \alpha_i(d) & \beta_i(-d) \\
    \beta_i(d)  & \alpha_i(-d)
  \end{array}
 \!\right)
\end{displaymath}
with the elements
%
%
 \begin{eqnarray}
 \alpha_i(d) &=& 
  \cos(k_id)\,+\,\frac{i}{2}\left(\zeta_i+\frac{1}{\zeta_i}\right)
  \sin(k_id) \\
 \beta_i(d)  &=& 
  \frac{i}{2}\left(\zeta_i-\frac{1}{\zeta_i}\right)\sin(k_id).
 \end{eqnarray}
The $\zeta_i$ are defined as
$\zeta_i=\mu_ik_0/(\mu_0k_i)$ or $\zeta_i=\varepsilon_0k_i/(\varepsilon_ik_0)$
for the TE and TM mode, respectively; 
indices zero refer to quantities in vacuum.
The transfer function coincides with 
the transmission coefficient $t$ (we choose $t_{-}$ for convenience)
of the imaging scattering matrix
$S_\mathrm{imaging}^{}=S\big[\tau_\mathrm{imaging}^{}\big]$,
\begin{equation}
t_{-} \ =\
\frac{
      \big[1+o(\delta)\big]\,e^{ik_0(a+b)}}
     {
      \cos(k_2d)+
      \Big[ 1+ \delta^2\Big(\frac{2\omega^4}{\omega^2-k_\|^2}\Big)
      \Big]\ i\sin(k_2d)
     }.
\end{equation}
We immediately recognize that the surface correction
$\delta^2\omega^4/(\omega^2-k_\|^2)=\delta^2\omega^4/k_2^2$ in the denominator
acts like an imaginary part
in the permeability or permittivity of the near perfect lens.
If the perfect lens condition $a+b=d$ is satisfied we have
$t_{-} \ =\ (1+o(\delta))/(1+\delta^2\omega^4 k_2^{-2}(1-e^{-2ik_2d}))$.
%
%
Let us now consider the transfer function for evanescent waves,
ie.\@ $k_\|>\omega$.
Then $k_2$ is purely imaginary such that the second term in the
denominator is always positive and the transfer function has no poles.
If $\delta^2\omega^4 k_2^{-2}(1-e^{-2ik_2d}) \ll 1$,
ie. for small $k_\|$,
we can neglect the $\delta$-correction in the denominator and find
an $o(1)$ behavior of the transmission function.
In the opposite limit of large $k_\|$ we can neglect the one in the
denominator and find the transfer function decaying exponentially
with $k_2d$.
The asymptotic exponential decay
$t_{-}\sim k_\|^2/(\delta^2\omega^4)\exp(-2k_\|d)$
can be used to define the cross-over momentum
$k_\mathrm{max}\,d\ =\ -\log\,(\omega^2\delta/k_\mathrm{max})$
%
%
which has an explicit solution in terms of the product-log function,
%
\begin{equation}
k_\mathrm{max}\,d
\ =\ -\mathrm{W}(-\omega^2\delta d) 
\ \sim\ -\log \delta
\end{equation}
for small $\delta$.
Since $\delta$ is assumed to be of the order of the discretization mesh 
constant this qualitatively represents the logarithmic dependence on the 
discretization observer in the TMM study above.


In conclusion
we investigated the transfer function of the discretized perfect lens by means
of FDTD and TMM simulations. 
The TMM has the advantage of computing the transfer function directly in 
$(k_\|,\omega)$-space as well as eliminating the problems associated with 
the explicit time dependence in the FDTD simulations.
We argue that the peak observed near $k_{\|,\mathrm{max}}$ in the FDTD 
transfer function is due to finite time artifacts; 
it does not exist in the TMM simulations.
Further we found that the finite discretization mesh acts like an imaginary 
deviations from the $\mu=\varepsilon=-1$ of the PL and leads to a cross-over
in the transfer function from $o(1)$ to exponential decay around a maximum
parallel momentum $k_{\|,\mathrm{max}}$ limiting the attainable 
super-resolution of the PL. 
We propose a simple qualitative model to describe the impact of the 
discretization in terms of effective thin $\mu$-only and $\varepsilon$-only 
surface layers exposed by the discretized LH slab which have a thickness 
$\delta$ that is of the order of the discretization mesh constant.
$k_{\|,\mathrm{max}} d$ is found to depend logarithmically on the mesh constant
in qualitative agreement with the TMM simulations.
Since virtually all simulation solve discretized
Maxwell equations they are all subject to this restriction.

This work was partially supported by Ames Laboratory 
(Contract number W-7405-Eng-82). 
Financial support of EU$\underline{~~}$FET project DALHM 
and DARPA (Contract number  MDA972-01-2-0016)
are also acknowledged.

\bibliographystyle{apsrev}

\end{document}